\documentclass[10pt,prl,aps,twocolumn,showpacs]{revtex4}
\usepackage{graphicx}

\usepackage{amssymb}

\begin{document}

\title{Global Phase Diagram of the Extended Kitaev-Heisenberg Model on Honeycomb Lattice}

\author{Jie Lou}
\affiliation{Department of Physics and State Key Laboratory of Surface Physics, Fudan University, Shanghai 200433, China}
\affiliation{Collaborative Innovation Center of Advanced Microstructures, Nanjing 210093, China}
\author{Long Liang}
\affiliation{Department of Physics and State Key Laboratory of Surface Physics, Fudan University, Shanghai 200433, China}
\affiliation{Collaborative Innovation Center of Advanced Microstructures, Nanjing 210093, China}
\author{Yue Yu}
\affiliation{Department of Physics and State Key Laboratory of Surface Physics, Fudan University, Shanghai 200433, China}
\affiliation{Collaborative Innovation Center of Advanced Microstructures, Nanjing 210093, China}
\author{Yan Chen}
\affiliation{Department of Physics and State Key Laboratory of Surface Physics, Fudan University, Shanghai 200433, China}
\affiliation{Collaborative Innovation Center of Advanced Microstructures, Nanjing 210093, China}

\begin{abstract}
We study  the extended Kitaev-Heisenberg (EKH) quantum spin model by adding bond-dependent off-diagonal Heisenberg term into the original KH model, which was recently proposed to describe the honeycomb Iridates.
A rigorous mathematical mapping of spin operators reveals the intrinsic symmetry of the model Hamiltonian.
By employing an unbiased numerical
entanglement renormalization method based on tensor network ansatz, we obtain the global phase diagram containing eight
distinct quantum phases. By using the dual mapping of spin operators, each of the individual magnetic phase in the global phase diagram can be clearly understood. At last, we show that a valence solid state emerges as the ground state in the quadro-critical region where multiple magnetic phases compete most intensively.

\end{abstract}

\date{\today}


\maketitle

The family of layered honeycomb iridium oxides Na$_2$IrO$_3$ and Li$_2$IrO$_3$  have attracted great interest both experimentally and theoretically~\cite{liu11,ye12,choi12,Jiang11,Reuther11,Trousselet11,Schaffer12,price12,singh10,singh12,subhro12,kimchi11,jackeli13,katukuri14,gret13,rau14,perkins14}
due to its possible relevance of Kitaev spin liquid~\cite{kitaev06}. Earlier theoretical studies proposed that the exchange from direct \emph{d}-orbital overlap of Ir ions may result in a Kitaev-Heisenberg (KH) model~\cite{jackeli10}. Extensive study of such model has shown a variety of fascinating phenomena, including an extended spin liquid phase and quantum phase transitions into several well-understood magnetic ground states. Experimentally, neutron scattering data revealed the presence of
the zigzag phase in Na$_2$IrO$_3$~\cite{liu11,ye12,choi12}. However, it appears to be hard stabilize within the
HK model; one may consider additional exchange paths~\cite{jackeli13} or further neighboring hoppings~\cite{kimchi11}.
Recently, a generic bond-dependent off-diagonal exchange term is included in the extended Kiteav-Heisenberg (EKH) model~\cite{rau14}.
By using a combination of classical techniques and exact diagonalization, 120$^{\circ}$ and incommensurate spiral order,
as well as extended regions of zigzag and stripy order are obtained.
This EKH model can be regarded as the minimal model to describe the essential physics of honeycomb iridium oxides.
However, the effect of bond-dependent off-diagonal exchange term is unclear and the obtained phase diagram is not clearly understood.

In this letter, we perform a systematic study of the EKH quantum spin model on honeycomb lattice by using the multiscale entanglement renormalization ansatz (MERA)~\cite{MERA} . We obtained a global phase diagram containing eight distinct quantum phases. A rigorous mathematical mapping of spin operators reveals the intrinsic symmetry of the model Hamiltonian. In particular, by rotation of spin operators obeying certain rules,
the additional bond-dependent off-diagonal Heisenberg exchange term can be mapped to Heisenberg or Kitaev interactions and vice versa. So the original EKH model can be mapped to the model of same form but with modified coupling constants.
By using this dual transformation, each of the individual magnetic phase in the
global phase diagram can be clearly understood. For instance,  some seemingly complicated magnetic orders can be mapped into simple antiferromagnetic (AFM) order or ferromagnetic (FM) order. Furthermore, both the finite size effect and the results for infinite system are discussed.
At last, we show that a valence solid state emerges as the ground state in the quadro-critical region where multiple magnetic phases compete most intensively.

\begin{figure}
\centerline{\includegraphics[angle=0,width=8.0cm]{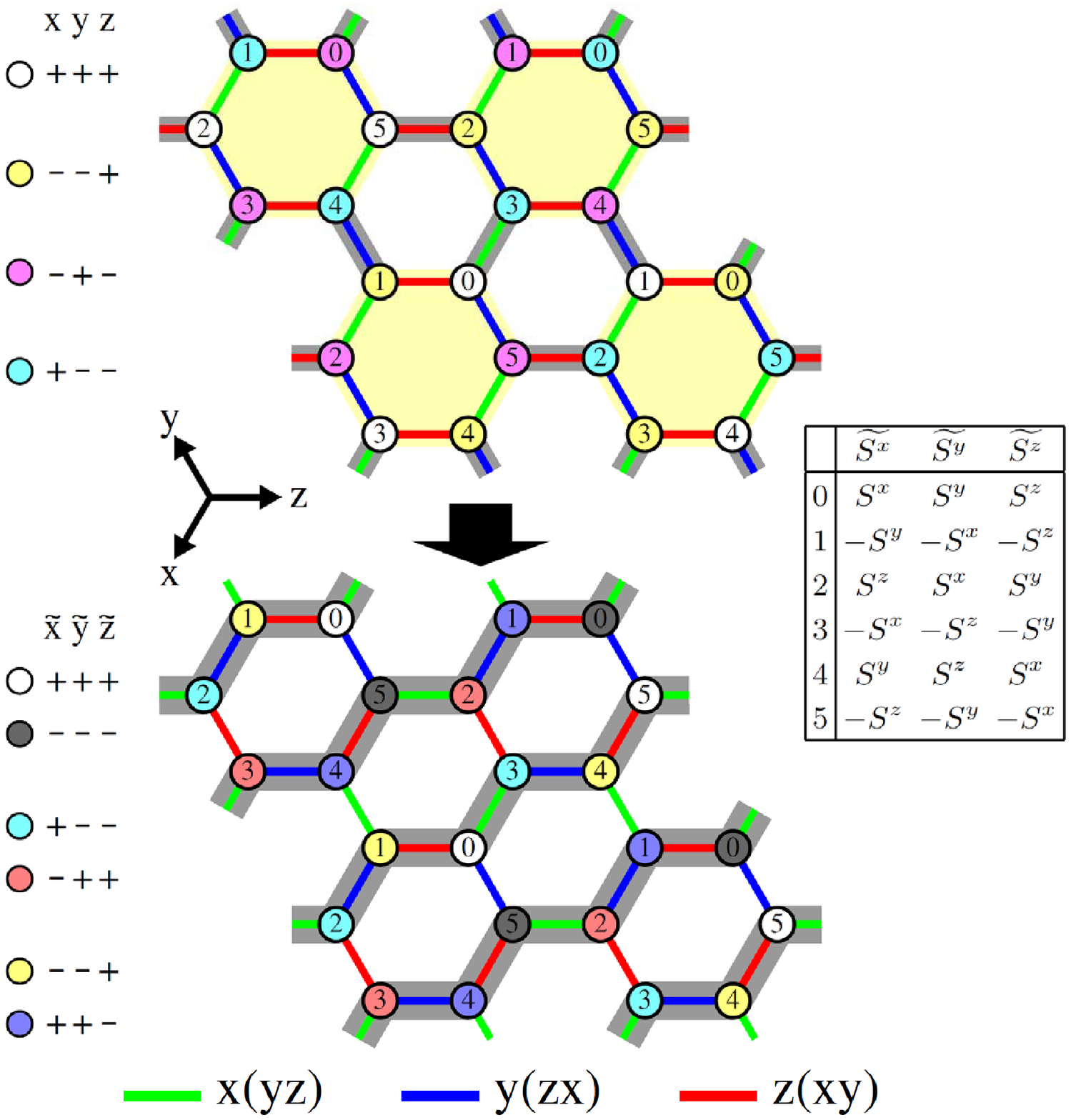}}
\caption{(color online). Mapping the original EKH model to its counterpart through spin rotations.
The upper panel stands for the original spin model, where spin directions
$\gamma(\alpha,\beta)$ for Kitaev and bond-dependent ($\Gamma$) exchanges
are denoted by different colors.
By performing the spin rotation marked by numbers 0 to 5, following the rule listed in
the table, we can map the original spin model to its counterpart.
Different colors are painted for each site to denote rules of sign change,
following which ZZ/SP phases can be mapped into AFM/FM.
Shades in the background in the upper panel represent the MERA structure we adopted in numerical calculation.
}
\label{illustration}
\end{figure}

\emph{Model Hamiltonian.}--- Our model is the EKH model defined on a honeycomb lattice, for which the Hamiltonian is
\begin{eqnarray}
H =\sum_{\langle i,j\rangle} [ J{\bf S}_i \cdot {\bf S}_j + KS_i^{\gamma}S_j^{\gamma} + \Gamma (S_i^{\alpha}
S_j^{\beta} + S_i^{\beta}S_j^{\alpha})],
\end{eqnarray}
where $\langle i,j \rangle$ denotes the nearest-neighbor, and terms with strength $J$, $K$ and $\Gamma$ represent Heisenberg, Kitaev
and bond dependent off-diagonal exchange interactions, respectively. We use $\gamma$ to distinguish one spin direction on each bond for the Kitaev exchange, as shown in Fig.~\ref{illustration}.
We have parameterized the energy scales so that $J^2+K^2+\Gamma^2=R^2$, $J = R\cos \theta \sin \phi$,$K = R\cos \theta \cos \phi$,
$\Gamma = R\sin \theta$. $\phi$ controls the ratio between Kitaev and Heisenberg terms.

It is well known that at the Kitaev limit $J=\Gamma=0, K \neq 0$,
the ground state of system is a topological spin liquid which has no long range order.
Away from the limit, several long range magnetic orders may emerge and compete with each other.
In addition to the AFM and FM order, the system also harbors stripy (ST) and zigzag (ZZ) phases,
in region centered at two points $K=\pm 2J$.
It can be shown that exactly at these two points,
the original KH hamiltonian can be mapped into a pure Heisenberg model with the coupling constant $J'=-J$.
The trick is to divide the system into four sublattices,
each flip spin components in accordance to certain rules, as denoted by different colors in Fig.\ref{illustration}.
In that sense, the ground states of KH model can be well understood in terms of the competitions among simple magnetic long-range orders.

\emph{Mapping of Spin Operators.}---In order to better understand the effect of $\Gamma$-term,
we applied rotation of the spin operators ${\bf S}^{\gamma} \rightarrow {\bf S}^{\alpha}$ (See the table in Fig.1) so that
the $\Gamma$ term can be mapped back to Kitaev exchange term.
By doing so, the original hamiltonian is mapped to the model of same form but with modified coupling constants,
\begin{eqnarray}
H &=&\sum_{<ij>}J'\widetilde{\bf {S}}_i \cdot \widetilde{\bf S}_j + K'\widetilde{S_i^{\gamma}}\widetilde{S_j^{\gamma}} + \Gamma ' (\widetilde{S_i^{\alpha}}
\widetilde{S_j^{\beta}} + \widetilde{S_i^{\beta}}\widetilde{S_j^{\alpha}}),
\end{eqnarray}
where strength of each term is related to original parameters as
$J'=-\Gamma$, $K'=\-(K-\Gamma-J)$, and $\Gamma'=-J$.
Notice that the Heisenberg and bond-dependent exchange couplings
are swapped through the mapping.

\begin{figure}
\centerline{\includegraphics[angle=0,width=9.0cm]{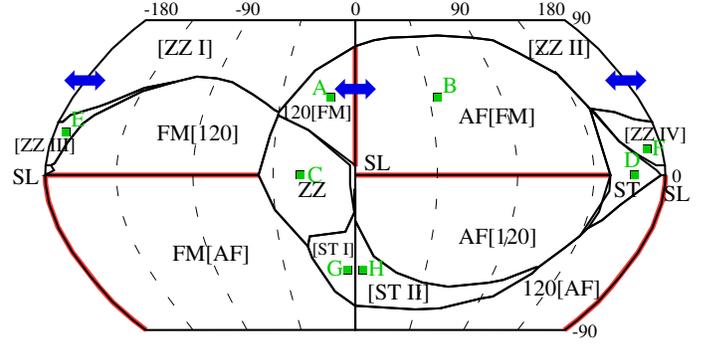}}
\caption{Global phase diagram from MERA calculations of 24-sites system.
Longitude $\phi$ and latitude $\theta$ of the world map parameterize the couplings of $J$, $K$ and $\Gamma$ terms.
In the upper panel, phases are described from the perspective of the original spin model,
whereas the lower panel shows the phase map when the rotated spin model is considered.
Regions of same type of magnetic order are shown in same color, labeled in the bottom.
Diamonds notates special points where the original model can be simplified to
simple Heisenberg interactions.
Red squares mark parameter sets for which we show spin configurations
of ground states obtained from ER calculations.
Phase boundaries marked by red wide lines corresponds to first order phase transition.}
\label{phase_map}
\end{figure}

\begin{figure}
\centerline{\includegraphics[angle=0,width=8.0cm]{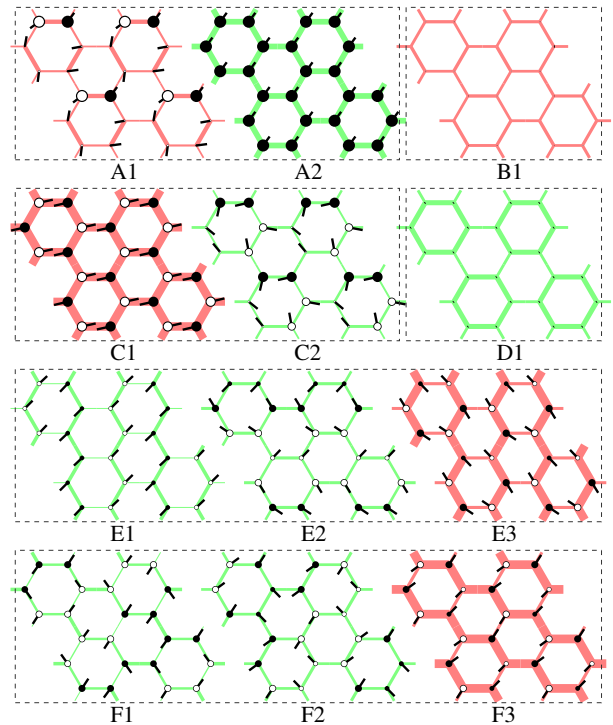}}
\caption{Spin configurations and nearest neighbors' correlations $\langle {\bf S}_i \cdot {\bf S}_j \rangle$
obtained from numerical calculation for 24-sites system.
On each site, spin $\langle {\bf S}_i \rangle$ is plotted as combination of a
circle (proportional to $\langle S_i^z \rangle$) and a vector (proportional to $\langle S_i^{xy} \rangle$).
Spin correlations are notated by bonds connecting spin pairs,
whose width are proportional to the strength of $\langle {\bf S}_i \cdot {\bf S}_j \rangle$.
Red/green color stands for FM/AF correlation, respectively.
Panel A $\sim$ F show results achieved in different phases, their corresponding parameter sets
can be read from the marks on phase maps.
The number of each notation stands for the perspective from which the ground state is viewed:
``1'' notates the original spin picture, ``2'' stands for the rotated spin picture,
and ``3'' means the result of zigzag mapping (for the rotated spin model),
described in Fig.~\ref{illustration}.
Results for larger systems can be found in Sup. material. }
\label{spinconfig}
\end{figure}

The mapping can be extremely helpful to understand the characteristics of ground state.
For instance, we take the limit $J=0, \Gamma=K=1$. Under rotations of spin operators,
the mapped Hamiltonian corresponds to a simple Heisenberg model with FM interaction $H=-\sum_{\langle i, j \rangle} {\bf S}_i {\bf S}_j$.
Indeed, numerical calculation confirms that its ground state is classical
FM state with energy per-site exactly equal to $-0.375$.
In Fig.~\ref{spinconfig}, panel (A), we show details of the ground state,
from perspectives before and after rotations of spins, for comparison.
In terms of original representation, the spin configuration of ground state is rather complicated, and long range magnetic order
can not be easily distinguished.
On the other hand, in the mapped representation, the spin configuration of ground state is a simple ferromagnet with
spins aligning in same direction.
Hence, the intrinsic symmetry of the model Hamiltonian may help us to better understand
the seemingly complicated magnetic orders of the system induced by the additional off-diagonal exchange term.
There exists several such parameter sets where the original model can be
simplified to pure Heisenberg models (shown in Fig.~\ref{phase_map}).
These special cases offer some additional hints for solving EKH model,
but it becomes rather complicated when various magnetic orders competes with each other, especially when the $\Gamma$ term dominates.

\emph{Global phase diagram.}---We use an unbiased MERA algorithm
to fully explore the phase diagram of EKH model on honeycomb lattice.
In the upper panel of Fig.\ref{illustration},
we show the ER tensor network we construct for the honeycomb lattice.
This method is capable of capturing the Kitaev spin liquid and other
phases where quantum fluctuations play an important role.
The system sizes we calculated range from 24 to 96.

The global phase diagram we numerically obtained is shown in Fig.~\ref{phase_map}, which is presented in format of a world map as functions of
two parameters $\theta$ and $\phi$. The antiferromegnet Kitaev point (spin liquid) is located exactly  at the map center,
and the original Kitaev Heisenberg model with $\Gamma=0$ corresponds to the equator line.
In Fig.~\ref{phase_map}, we present two phase maps and each of them belongs to a different spin representation.
The very same state show different magnetic long range orders without (upper panel)
or with (lower panel) spin rotations. Here we denote these two different orders with labeling
``X'' and ``$\bar{Y}$''.
By performing the spin rotations, we notice that Heisenberg and $\Gamma$ exchanges
are swapped, namely, $\widetilde{J}=-\Gamma$ and
vice versa.

\emph{AFM/FM and 120$^{\circ}$ phases}---Due to the fact that the EKH model is mapped to itself,
we expect the phase map to be self consistent by spin rotations.
Such mapping correspondence is clearly observed for AFM and FM orders,
which occupy a large portion of the phase map.
It is important to notice that one block of certain order (AFM for example) in the original
map can be further separated into sections of different orders ($\widetilde{120}$ and $\widetilde{FM}$)
in the phase map for rotated spins, vice versa.

This feature seems to be contradictory, but careful analysis reveals that these two regions are indeed of different magnetic orders.
As a matter of fact, a first order phase transition is clearly shown
when the phase boundary is approached and eventually crossed (see Figure in supplementary material).
The physical reason can be understood as the following.
For the $\widetilde{FM}$ order, spins $\langle \widetilde{{\bf S}}_i \rangle$ are parallels for all sites.
Naturally, we expect equal spin components for every site, namely $\widetilde{S_i^x}=\widetilde{S_j^x}$.
On the other hand, the very phase is of the AFM order in term of original spins,
which require $|S_i^x|=|S_j^x|$.
Notice that rotations of spins ($S_i^x \rightarrow -\widetilde{S_i^y}$, for example) are involved
when original hamiltonian is mapped to the new one.
As a result, the only possible outcome for these two requirements to be met at the same time
is that all spins order along $(1,1,1)$ direction, namely $S_i^x=S_i^y=S_i^z$.
In another word, the bulk of $\widetilde{FM}/AF$ phase is of Ising type.
This is to be expected, as introduction of $\Gamma$ term breaks SU(2) symmetry of
original Heisenberg magnets.

Similar situation occurs for the $\widetilde{120^{\circ}}/AFM$ phase.
The $\widetilde{120^{\circ}}$ phase is named after the fact that all angles between
next-nearest neighbor spins $\widetilde{S}$ are $2\pi/3$ in this order.
an illustrated in Fig.~3.
Such phase appear south of euqator, which represent a positive nearest neighbor Heisenberg couplings
$\widetilde{J}=-\Gamma$, unfavoring the $\widetilde{FM}$ phase.
Therefore, spins in original picture orders in the plane perpendicular to $(1,1,1)$, as a compromise.
(Notice, the system is still in the AFM phase with all nearest neighbors spins antiparallel.)
It can be shown that whence $S_i^x+S_i^y+S_i^z=0$ (true for any vector in plane
perpendicular to $(1,1,1)$), after rotations listed in tables,
the $\widetilde{120^{\circ}}$ order emergies.
In this case, the U(1) symmetry, instead of SU(2) symmetry, is broken in the AFM order.

Due to effect of $\Gamma$ term, the orginal SU(2) symmetric Neel AFM order breaks
into two different types: an Ising AFM north of the equator, and a U(1) AFM south.
A first order phase transition shows up at the phase boundary.
They are different physical orders with distinct broken symmetry.
Such phenomena is universal for all AFM and FM orders, in both original and rotated spin models.
The finite size effect is less remarkable as shown in our numerical calculation.
All magnetic orders remain the same, except a slightly shifts of phase boundaries \cite{Supmat}.

\emph{Zigzag/stripy phases and frustration}---While switching longitude and latitude is helpful to understand
the relation between two phase maps, especially for AFM/FM orders, it does not work well when zigzag/tripy phases are considered. As a matter of fact, these phases are far more complicated than Neel-like orders discussed above.
From numerical calculation, we find that spins are much more prone to tilting away than
strictly order along one direction.
Meanwhile, various candidates of magnetic orders compete for the ground state,
and energy gaps between them are relatively small.
The reason for such different behaviors can be found in the mapping
when rotations of spins are done.
As mentioned before, such discrepancy may not affect AFM and FM orders much,
(as we observed in our numerical calculation), but do change zigzag phase and stripy phase
significantly.
As a matter of fact, the new pattern of Kitaev and $\Gamma$ interactions is frustrated
in terms of $\widetilde{zigzag}$/$\widetilde{stripy}$ order.

For the original Kitaev Heisenberg model,
a four-sublattice mapping between the ST/ZZ order
and the AFM/FM order is well known and discussed~\cite{jackeli10}.
Naturally, we expect same type of operation can be applied for the rotated EKH model.
Unfortunately it is not possible to obtain such a perfect mapping in this case.
One can show that, when we flip spin operators one by one along an hexagon,
according to arrangement of Kitaev interaction $\gamma_{i,j}$ on each bond,
we can not obtain self-consistency after a full circle.
As a result, we have to sacrifice certain bonds in order to map the rotated spin model
to its $\widetilde{ZZ}$/$\widetilde{ST}$ counterpart.
One way to make the compromise is shown in the lower panel of Fig.~\ref{illustration}.
We keep the sign-flipping rule along chains following certain direction,
and reverse spin operators when jump between chains.
Clearly there exists 3 degenerate ways(directions) for such sign change rules.
Only one is plotted in the figure, and shades indicate the chosen direction of
$\widetilde{ZZ}$/$\widetilde{ST}$ chains.

Such a compromise sacrifices certain Kitaev interactions intra-chain.
Numerical study shows that such a mapping is energy-favorable for most
zigzag region in the rotated spin model.
Examples of such $\widetilde{ZZ}$ phases are shown in Fig.~\ref{spinconfig}.
In the rotated spin model, we expect the $\widetilde{ZZ}$ phase to appear near the point
$\widetilde{K}=-2\widetilde{J}, \widetilde{\Gamma}=0$, which is translated to
$J=0, K=-1, \Gamma=1$.
Notice that the spin-spin correlation shows that intra-chains' correlations are significantly
larger than inter-chains', which is consistent with the mapping we made.
It is necessary to mention that due to this frustration,
these states become less stable when we tune parameters away from the zigzag point.
spin bend away and new magnetic order appears.
One notice that for $J<0$ and $J>0$, the $\widetilde{zigzag}$ phase for rotated spin model also
emerges as different magnetic order in original spin picture.
The reason is very similar to the AF case explained above.
Due to perturbation of $\widetilde{\Gamma}=-J$ term, system magnetize is different
quadrant in these two cases. Details can be found in \cite{Supmat}.

The mapping we choose is not necessarily the unique one, however, especially in larger systems.
In small system with 24 and 36 sites, numerical calculations shows that
the former mentioned states are the ground states spanning the zigzag
region for the rotated spin model.
However for larger systems, various magnetic order candidates contradict to such mapping
also appears, with reasonably good energy~\cite{Supmat}.
With current calculation capability, we are not able to probe the region thoroughly.

\begin{figure}
\centerline{\includegraphics[angle=0,width=8.0cm]{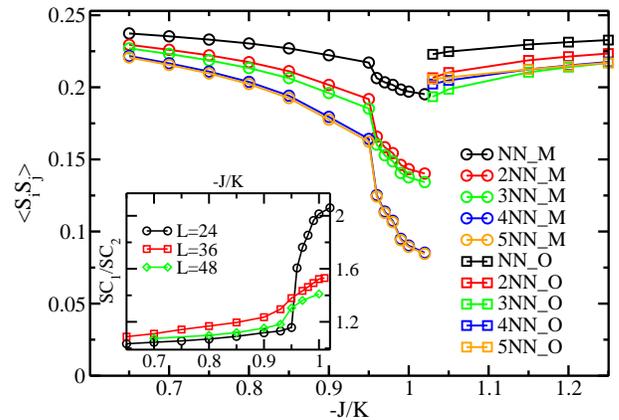}}
\caption{
Evolution of spin correlations (SC) $\langle {\bf S}_i \cdot {\bf S}_j \rangle$ when
system undergoes a transition from magnetic long range orders to the VBS phase.
$K$ and $\Gamma$ are fixed to be $1$, and $J$ is tuned through the phase transition.
Different colors represents various distances between two spins $i$ and $j$,
from nearest neighbor (NN) to 5th nearest neighbor (5NN), the maximal distance in a
periodic 24-site lattice.
At $-J/K=0.95$, a phase transition between $\widetilde{FM}$ and the VBS order
occurs, and correlations for rotated spins (SC notated by suffix M) drops abruptly.
When $J/K$ goes over $1$, the system enter the FM order, and SCs for original spins
(suffix O) do not decay by increasing the separation.
The inset shows the ratio between average spin correlation intra and inter a plaquette
hexagon $SC_1/SC_2$. Results for various system sizes are plotted.
}
\label{VBS}
\end{figure}

\emph{Valence bond solid.} --- As expected, multiple magnetic long range orders compete most intensively
close to two points $-J=K=\Gamma=1$ and $J=-K=-\Gamma=1$.
These two special points corresponds to verging points of multiple phases,
as shown in Fig.~\ref{phase_map}.
It is not clear what is to be expected from analytical analysis,
and our numerical calculation shows that small regions surround these
multi-critical points harbor the VBS phase.
A periodic strong-weak shifting pattern for nearest neighbor spin correlation
can be observed in Fig.~\ref{spinconfig}, panel D2.
More precisely speaking, correlations inside a plaquette (hexagon unitcell) is larger than
intra-plaquette ones.
As a result, we expect long range spin correlations to drop fast when the
distance between two separated spins is increased.
In contrast, for all magnetic orders
discussed above, correlations decay much slower when spins are further separated.
In Fig.~\ref{VBS}, we show such results from numerical calculations and a transition
between the magnetically ordered phase (120/$\widetilde{FM}$) and the VBS phase can be distinguished.

To further clarify this, we define spin correlations intra/inter plaquette
hexagon (P) as $SC_1$ and $SC_2$, where $SC_1 = \sum_{P_i=P_j} {\bf S}_i \cdot {\bf S}_j$ and $SC_2 = \sum_{P_i \neq P_j} {\bf S}_i \cdot {\bf S}_j$.
In subset of Fig.~\ref{VBS}, we show that for various system sizes we calculated,
the ratio between $SC_1$ and $SC_2$ increase significantly when the VBS
region is approached.
It is worth mentioning that the local spin expectation value $\langle {\bf S}_i \rangle$
vanishes in these critical regions (as shown in Fig.~\ref{spinconfig} for 24-sites system),
a similar feature observed for the Kitaev spin liquid.
However, long range spin correlations reveal that they are intrinsically different.
We have to clarify that our calculation is limited to relatively small system sizes,
as a result, finite size effect boundary conditions may play important roles.
We can not rule out the possibility of spin liquid in this critical region.

\emph{Conclusions.} ---We numerically studied the global phase diagram of the EKH model containing eight
distinct quantum phases. A rigorous mathematical mapping of spin operators reveals the intrinsic symmetry of the model Hamiltonian.
 By using the dual mapping of spin operators, each of the individual magnetic phase in the phase diagram can be clearly understood.
 Finally, we showed that a valence solid state emerges as the ground state in the critical region where multiple magnetic phases compete most intensively.

\emph{Acknowledgments.}--- This work was supported by the State Key Programs of China (Grant No. 2012CB921604),
the National Natural Science Foundation of China (Grant Nos. 11274069, and 11474064).

\end{document}